\documentclass[global,twocolumn]{svjour}

\newcommand{\be}{\begin{eqnarray}}
\newcommand{\ee}{\end{eqnarray}}

\newcommand{\Shalf}{$^{2}$S$_{1/2}$ }
\newcommand{\Szero}{$^1$S$_0$ }
\newcommand{\Phalf}{$^{2}$P$_{1/2}$ }
\newcommand{\Fstate}{$^{2}$F$_{7/2}$ }
\newcommand{\Pone}{$^1$P$_1$ }
\newcommand{\Dhalf}{$^{2}$D$_{3/2}$ }
\newcommand{\jK}{$^{3}[3/2]_{1/2}$ }

\newcommand{\ybg}{\textsuperscript{172}Yb}

\newcommand{\ybuion}{\textsuperscript{171}Yb\textsuperscript{+}}

\newcommand{\Vpp}{V$\!_{\rm{pp}}$}
\usepackage{textcomp}
\usepackage[tight]{units}
\usepackage{graphicx}
\usepackage{subfigure}
\usepackage{color}

\begin{document}

\title{Thick-film technology for ultra high vacuum interfaces\\ of micro-structured traps}

\author{Delia~Kaufmann, Thomas~Collath, M.~Tanveer~Baig, Peter~Kaufmann, Eman~Asenwar, Michael~Johanning,  Christof~Wunderlich}

\institute{Faculty IV: Science and Technology, Department of
Physics, University of Siegen, 57068 Siegen, Germany \\
\email{wunderlich@physik.uni-siegen.de}}

\titlerunning{Thick-film technology for ultra high vacuum interfaces of micro-structured traps}
\authorrunning{D.~Kaufmann et. al.}
\date{Received: date / Revised version: date}
\maketitle

\begin{abstract}
We adopt thick-film technology to produce ultra high vacuum
compatible interfaces for electrical signals. These interfaces
permit voltages of hundreds of volts and currents of several amperes
and allow for very compact vacuum setups, useful in quantum optics
in general, and in particular for quantum information science using
miniaturized traps for ions \cite{Kielpinski2002} or neutral atoms
\cite{Schmiedmayer2000,Treutlein2006,Schmiedmayer2006}. Such printed
circuits can also be useful as pure in-vacuum devices. We
demonstrate a specific interface, which provides eleven current
feedthroughs, more than 70 dc feed\-throughs and a feedthrough for
radio frequencies. We achieve a pressure in the low \unit[$10^{-11}$]{mbar} range and
demonstrate the full functionality of the interface by trapping
chains of cold ytterbium ions, which requires the presence of all of
the above mentioned signals. In order to supply precise
time-dependent voltages to the ion trap, a versatile multi-channel
device has been developed.
\end{abstract}



\maketitle



\section{Introduction}
\label{sec:Introduction}

Many experiments in atomic and molecular physics are conducted under
ultra high vacuum (UHV) conditions to obtain low collision rates
with background gas. At the same time, many of these experiments,
especially those related to quantum information science
 require versatile and detailed control
\cite{Treutlein2006,Schmiedmayer2006,Rowe2002,Stick2006,Schulz2008,Harlander2011,Pearson2006,Allcock2010,Moehring2011,McLoughlin2011,Brownnutt2006}. For a recent review of micro-structured ion traps, see \cite{Hughes2011}.
Unless control is applied using optical fields only \cite{Bakr2010,Weitenberg2011},
control is usually required in terms of a large number of
control voltages
\cite{Kielpinski2002,Rowe2002,Stick2006,Schulz2008,Harlander2011,Pearson2006,Allcock2010,Moehring2011,McLoughlin2011,Brownnutt2006}
or currents
\cite{Schmiedmayer2000,Treutlein2006}, and rf or microwave fields
\cite{Schmiedmayer2006,Boehi2010,Karin2011}.

\begin{figure}
\centering
\includegraphics[height=6cm]{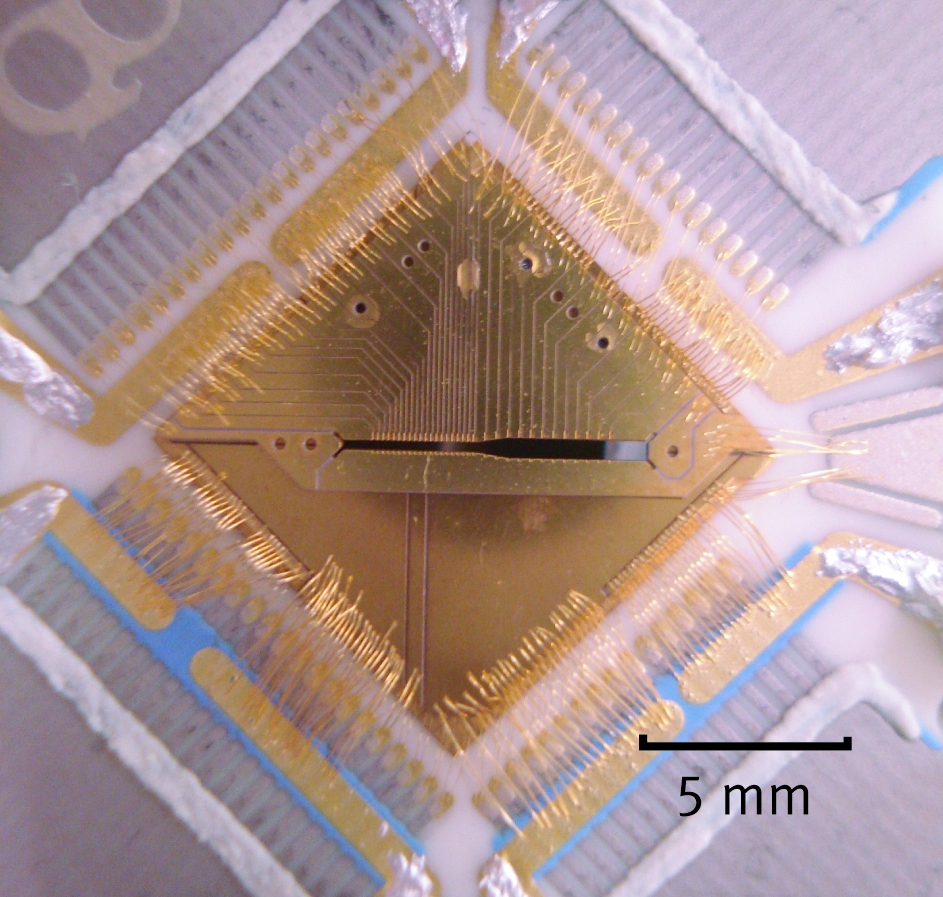}
\caption{The wire bonded microtrap chip (the trap itself was developed in collaboration with the group of F.~Schmidt-Kaler \cite{Schulz2008,Schulz2006}).  The wire bonds provide dc and rf connectivity, a large
number of wire bonds is dedicated to the current carrying middle layer.
Distributing the current over several wire bonds allows to apply large currents.
}\label{fig:BondedChip}
\end{figure}

Scaling down such experiments is often desirable, either to improve
or simplify trapping, to increase the level of detail on the control
of the trapping potential or to make them interesting and
competitive for industrial applications. In this case commercially
available interfaces impose a serious limit for miniaturization. As
a solution, we introduce thick-film technology as a method to
produce custom made vacuum interfaces for a large number of signals
(as dc and rf voltages and currents) with a small footprint and show
that by sealing such parts against simple custom made vacuum
components UHV pressure can be attained.

\begin{figure*}
\centering
\includegraphics[width=2\columnwidth]{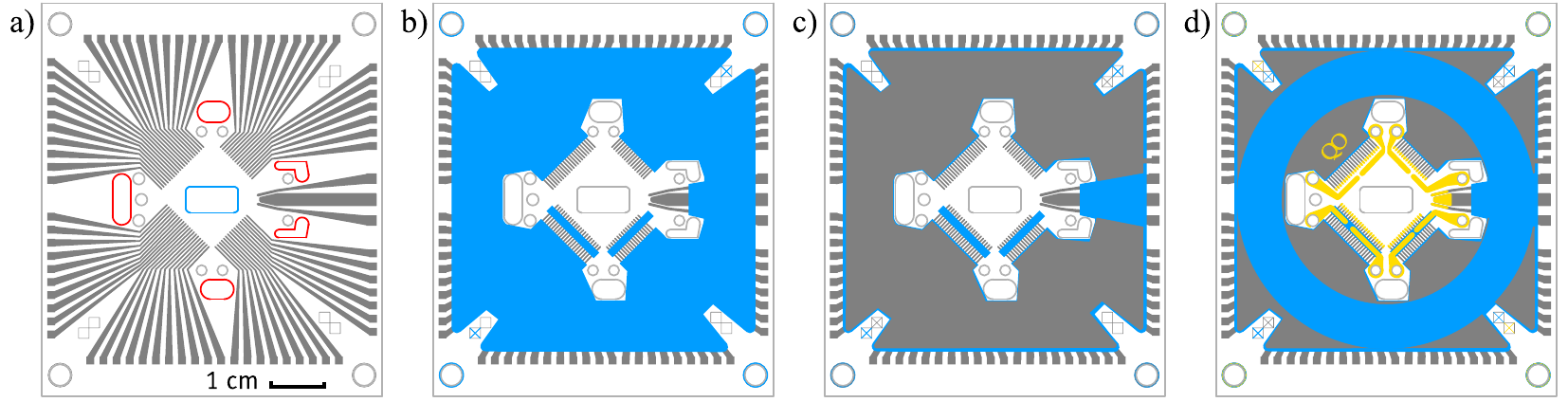}
\caption{Steps of thick-film printing: as an example we show the top
layer which provides rf and dc connectivity for the trap. The first
printing step (a) establishes the radial wires for dc (thin) and rf
(thick) connections, visible as grey wires. If filtering
 is not required, already one
additional isolation layer (printed in blue in step (b)) suffices to
complete the interface as this avoids shortcuts induced by indium
seals (see vacuum setup in section~\ref{sec:Vacuum}). The third step
(c) adds a large conductive area (again in dark grey) which acts as
a filter capacitor. In picture (d) two more prints have been added:
the circular blue ring of isolating paste separates the filter
capacitor from any indium seal, printed gold pads (displayed as
yellow) improve the adhesion for wire bonding. Also indicated in
figure (a) are holes for the trap (blue) and for pumping
(red).}\label{fig:printingSteps}
\end{figure*}

In the remainder of this article, we discuss thick-film technology
and its applicability to UHV setups and present the design and
describe the manufacturing steps of a specific interface. This
interface is used to operate a micro-structured segmented linear ion
trap for quantum information science experiments (see
figure~\ref{fig:BondedChip}). This trap was designed to trap and
laser cool tens of ions in a locally tunable axial potential, and it
allows the application
of a switchable magnetic field gradient to implement
\textbf{MA}gnetic \textbf{G}radient \textbf{I}nduced
\textbf{C}oupling (MAGIC)
\cite{Mintert2001,Mintert2001E,Wunderlich2002,Wunderlich2003,Ospelkaus2008}.
MAGIC provides coupling between internal and motional states
and establishes coupling of all spins, mediated by
the ions common vibrational motion. MAGIC recently received attention, as it allows to manipulate quantum information with rf \cite{Johanning2009b} and microwave fields and allows for fast laserless conditional gates \cite{Ospelkaus2011}. Furthermore microwave gates were demonstrated with high fidelity \cite{Brown2011},  robust states using microwave dressing fields were implemented \cite{Timoney2011}, as well as gates which are largely insensitive to detuning and amplitude errors \cite{Timoney2008}, the entanglement of non-neighboring ions was shown  \cite{Khromova2011},  and it has been discussed as a promising approach to implement quantum simulations \cite{Johanning2009,Welzel2011}.

\section{Micro-structured ion trap}\label{sec:SegmentedTrap}

Here, we describe the operation of a micro-struc\-tured segmented
linear ion trap. Such traps allow the tuning of trapping parameters globally and locally. They thus
facilitate the splitting and merging of ion strings to implement
scalable quantum computing based on simultaneous operations on only
two ions \cite{Kielpinski2002}, as well as the tailoring of
interaction Hamiltonians \cite{McHugh2005,Wunderlich_H2009}.
Providing this amount and detail of control, such traps require a
large number of electrical signals. Among these are rf signals with
amplitudes of hundreds of volts to achieve trapping of ions, a large
number of dc voltages (here, up to $\pm$10 volts) to shape the axial
trapping potentials, and currents up to a few amperes for atom ovens
and (in case of employing MAGIC) for the creation of magnetic fields
and gradients.

Our trap is based on a segmented 3-d linear trap design developed in
the group of F.~Schmidt-Kaler  \cite{Schulz2008,Schulz2006}, other three dimensional microstructured traps are
described, for instance, in \cite{Rowe2002,Stick2006,Harlander2011,Brownnutt2006}.
Ions are trapped in the slit of a linear Paul trap
composed of three laser-machined layers made of alumina
(Al$_2$O$_3$). The outer gold-coated layers (in the following called top  and bottom layer)
feature electrically separated areas which form a pattern of dc and rf electrodes that
provides the entire trapping functionality.
The trap offers a wide loading region with nine segments, a narrow
region for tight confinement with nineteen segments and a three segment taper to
connect the two regions (see figure~\ref{fig:BondedChip}).

The photo in figure~\ref{fig:BondedChip} shows the mounted
microtrap. The trapping potential is generated within the slit in
the middle of the quadratic trap chip structure. Inside the loading
region the slit is \unit[500]{\textmu m} wide and narrows down to
\unit[250]{\textmu m} in the narrower trapping region. The total
length of the slit is \unit[7.4]{mm}, the edge length of the whole
trap chip is \unit[11]{mm}. The micro-structured design of the trap
allows for trapping ions in different trap regions, moving them
along the slit and applying magnetic gradients to the trapped ions.

The middle layer (stacked between the outer layers) was specifically
developed to provide an adjustable magnetic field gradient. This
layer is made of gold coated, electroplated alumina. We use the
conductivity of this layer to apply a current for generating
magnetic fields and gradients in the proximity of the trapped ions,
in order to address and couple trapped ions in the MAGIC-scheme. To
boost the coupling induced by the magnetic field gradient, it is
desirable to attain as large gradients as possible and this design
was optimized for low dissipated heat per gradient, as the current
limit through the coil is given by the damage threshold of the coil
structure and its connections. In a preliminary test, we found that
several amperes could be applied
without any damage. As this preliminary test was done on a test
structure with high ohmic resistances and no cooling, we expect to
be able to apply more than \unit[10]{A} of current on the setup. To
avoid the wire bonds being a bottleneck, the current is
distributed over many bonding wires when feeding it to the chip. The
large number of additional wire bonds required cropping of the top
layer, to provide sufficient space for large bond pads on the middle
layer (see figure~\ref{fig:BondedChip}). Laser cut Polyimid foils
inserted between all trap layers isolate the entirely gold coated
middle layer from the outer trap layers electrically. Details on the
design and functionality of the magnetic field generating elements
will be published elsewhere. In order to provide all necessary
connections, a compact solution was developed and is described in
what follows.

\section{Microtrap Chip Carrier}
\label{sec:ChipCarrier}

\subsection{Thick-film technology}
\label{sec:ThickFilmTechnologyBasics}

We adopt thick-film technology to design a compact carrier for
micro-structured traps which simultaneously acts as a vacuum
interface. Thick-film technology is a well-known technique for the
fabrication of electronic circuits \cite{Harper1974,Gupta2005}. With
a screen printing process, wire structures are printed onto ceramics
or other suitable insulating materials. After the printing process,
e.g. on alumina, the printed conducting paste has to be fired at
\unit[850]{$^\circ$C}. Small glass particles inside the conducting paste fuse, and the
printed structures reliably cling to the ceramic surface. After the
first printing has been fired, more complex structures and circuits
can be generated by successively adding more layers on top of an
already fixed structure.

The paste used in the process is not limited to conducting
paste (which can contain e.g. silver and
palladium, or gold), but also isolating pastes as well as
resistor pastes and pastes to create specific dielectric layers (e. g. to
create printed capacitors) are available. Thick-film technology can be used
to produce wire structures with a minimum width of less than \unit[0.1]{mm}.
Using a laser to trim the printed circuits,
even smaller structures can be fabricated. With the same technique,
thick-film resistors can be adjusted to accuracies of better than
\unit[0.1]{\%}.
The wire bonding technique, conducting glue and even common
soldering can be used to connect the printed structures to certain
setups.

We found that thick-film printed ceramics can be used in UHV
environments (see Sec.~\ref{sec:Vacuum}). For our purpose, alumina is
advantageous as a base material, because the heat conductivity of this
ceramic is comparatively high (about \unit[25]{W/(K$\cdot$m)}) and allows to
remove any thermal intake from ohmic heating.

\begin{figure}
\centering
\subfigure[~]{\includegraphics[width=.45\columnwidth]{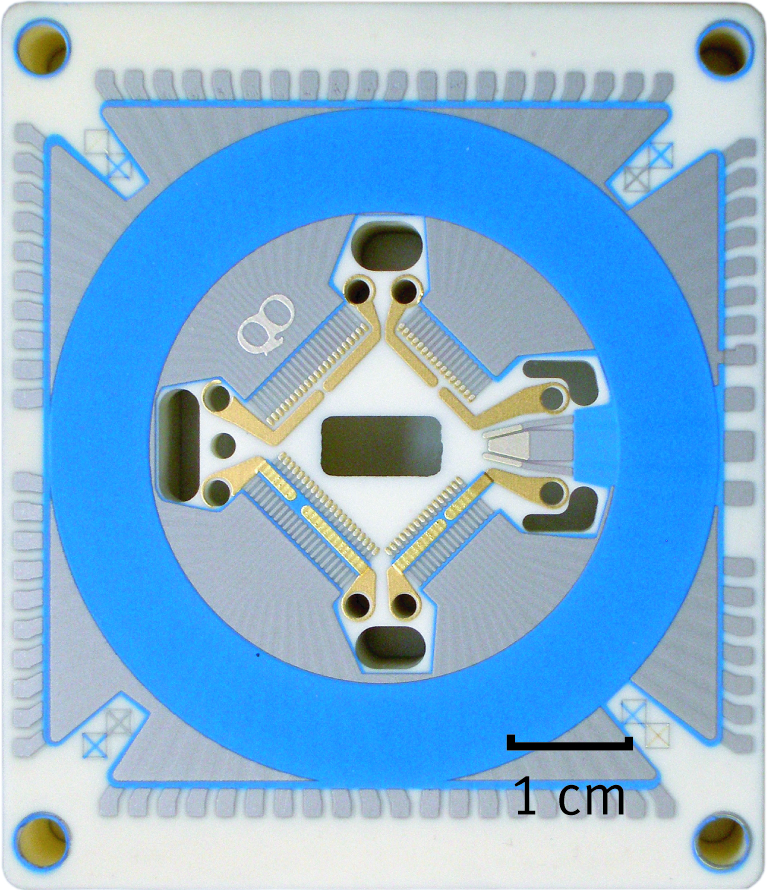}}
\subfigure[~]{\includegraphics[width=.45\columnwidth]{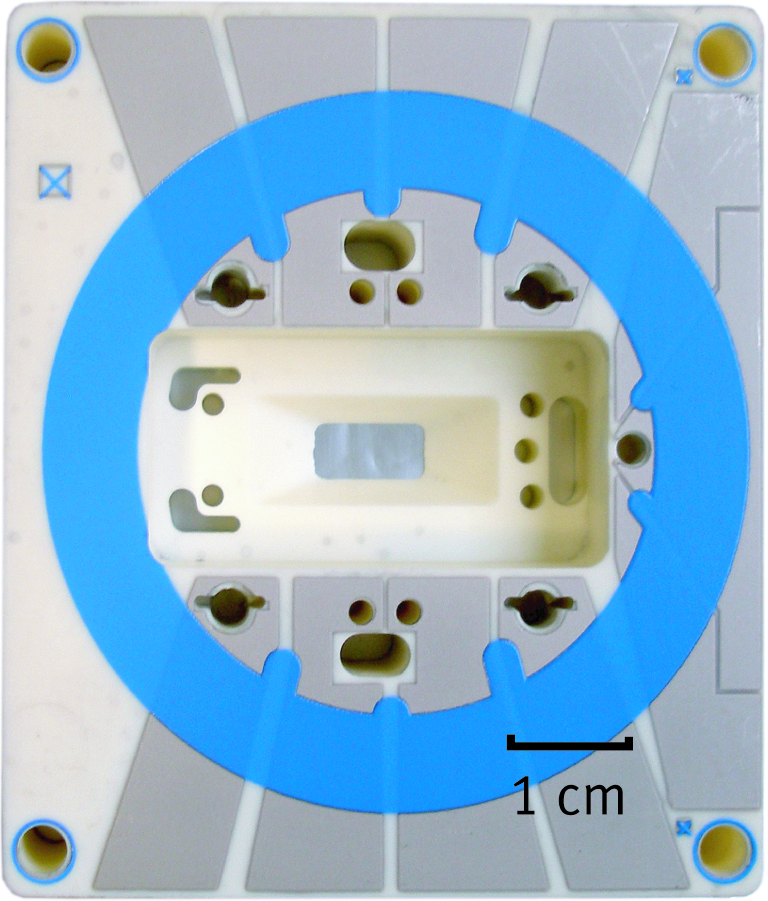}}
\caption{The microtrap chip carrier after all printing steps are
applied, before mounting and wire bonding of the trap chip (a) top
view: The thick-film printing procedure was detailed for this side
in figure~\ref{fig:printingSteps}. This side features a large number
of dc channels with a small footprint and the rf supply. (b) bottom
view: this side was optimized for currents (for ovens and solenoids)
and features wide conductors and large contact
areas.}\label{fig:Chipcarrier}
\end{figure}

\subsection{Design and implementation}
\label{sec:ChipCarrierRequirements}

In order to conduct experiments with the microtrap, a radio frequency voltage, 70 dc-voltages, and
currents of several amperes have to be applied to the trap.
Furthermore, heat generated from ohmic losses needs to be removed.
As ion trapping can only be performed
in an UHV environment, all signals have
to be fed through vacuum interfaces. In order to reduce noise and rf
crosstalk on dc electrodes, low pass filters connected to every
dc electrode should be located close to the trap (in our design we achieve less than \unit[10]{cm}).

\begin{figure*}
\centering
\includegraphics[width=13cm]{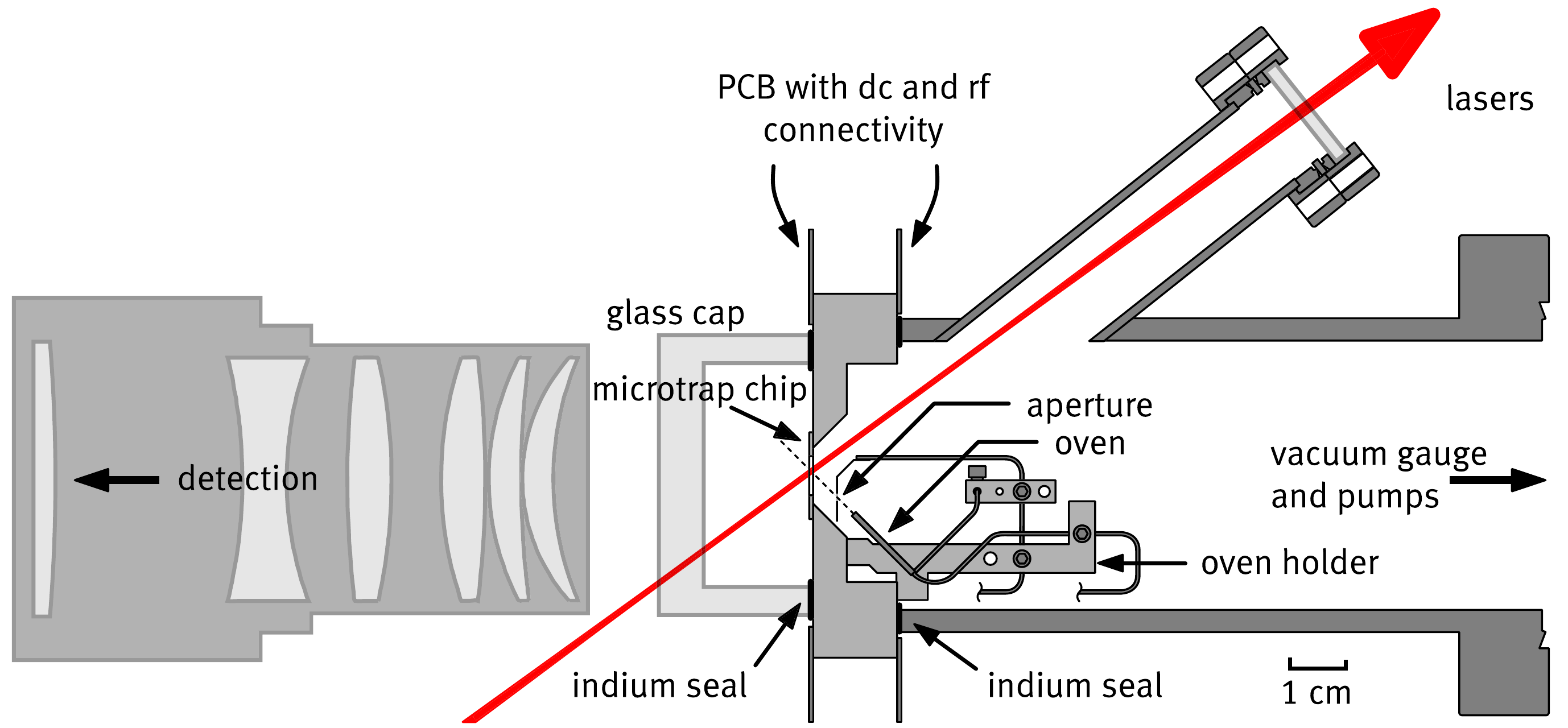}
\caption{Schematics of the chip carrier embedded into the vacuum and optical system,
as seen from top: fluorescence light is detected via light gathering
optics with a numerical aperture of 0.4 (left); the rest of the vacuum setup, containing an ion
getter pump, a titanium sublimation pump and an ion gauge (connected to the right)
contains no further feedthroughs for the microtrap.
}\label{fig:setupTopView}
\end{figure*}

All these requirements were implemented with a single thick-film
printed ceramic block 
forming the microtrap chip carrier. This chip carrier consists of a
\unit[15]{mm} thick block of alumina with outer dimensions of
\unit[65]{mm} $\times$ \unit[75]{mm}.

The design of the thick-film printed structure for the upper side of
the chip carrier is illustrated in figure~\ref{fig:printingSteps}.
From left to right, the sequence of the printing process is
indicated to allow a detailed view of the complex structure.
Figure~\ref{fig:printingSteps}a shows the blueprint of
silver-palladium wires connecting to the 70 dc electrodes, and the
tapered rf electrode's connection with two shielding wires printed
next to it. To fabricate low pass filters close to the microtrap,
the next two printing steps provide a capacitance overlaying all
wires, which connects to all dc electrodes. In order to form this
capacitor, blue isolation paste was printed upon the wire structure
in a first step (see figure~\ref{fig:printingSteps}b).
Figure~\ref{fig:printingSteps}c shows the structure of the
conducting capacitance layer. In the last two printing steps  (shown
in figure~\ref{fig:printingSteps}d), a ring of blue isolation paste
for the vacuum interface, and golden pads for wire bonding were
fabricated. The isolation layer of the capacitor was printed twice
in order to prevent shortcuts. For the ring, three isolating layers
provide a thickness of at least
\unit[20]{\textmu m} to apply polishing (see section
\ref{sec:Vacuum}). Although we gained positive experience with
resistive pastes, we keep all resistors for filtering outside vacuum
on nearby exchangable filter boards to keep them fully flexible (see
section~\ref{sec:TrappingPotentials}). The thickness of the chip
carrier is \unit[15]{mm} and thus provides good mechanical stability
and allows to add stud holes for heat pipes to cool the whole
carrier from the air side.

The microtrap chip carrier, completely fabricated, is displayed in
figure~\ref{fig:Chipcarrier}a and \ref{fig:Chipcarrier}b.
Figure~\ref{fig:Chipcarrier}a shows the top view of the carrier,
designed as illustrated in the blueprints of figure
\ref{fig:printingSteps}. This final thick-film printed design
provides a capacitance of \unit[(44~$\pm$~3)]{pF} for each dc
connecting wire. Several holes in the ceramic block (indicated in
red in figure~\ref{fig:printingSteps}a) are pumping holes to provide
good pumping through the carrier. The rectangular hole in the middle
of the carrier (indicated in blue in
figure~\ref{fig:printingSteps}a) is located where the trap is
mounted. All other holes are feedthroughs for screws in order to
connect electrically the upper and the lower side of the chip
carrier and to fix the atomic ovens (see overview in
figure~\ref{fig:setupTopView}). Figure~\ref{fig:Chipcarrier}b shows
the bottom view of the chip carrier, which is used to provide high
current supplies for the atomic ovens and the coil structure
providing the magnetic gradient. Furthermore, it contains a
thick-film printed isolation ring similar to the one on the top side
of the carrier.

The trap is glued to the carrier and electrically connected with wire bonds (see figure~\ref{fig:BondedChip}).
A reliable connection is ensured by connecting each dc electrode with two bonding wires.

\section{Vacuum}
\label{sec:Vacuum}

\subsection{Indium seals}
\label{sec:IndiumSeals}

To achieve a pressure in the UHV  range required for this experiment
we found that placing rings of indium between polished surfaces of
the vacuum components offers a reliable sealing method. The
essential vacuum system can be seen in
figure~\ref{fig:setupTopView}. The chip carrier is placed between a
glass cap with an interferometric quality window towards the left,
which ensures good optical access and a steel chamber which connects
to standard UHV components (pumps and a gauge) with ISO-CF flanges
to the right. The contact surfaces of the glass cap and the steel
chamber are polished, but no optical surface quality is required.
The counterparts of these surfaces on the chip carrier (the rings of
isolating paste, visible as blue rings in
figure~\ref{fig:Chipcarrier}) were polished as well.

Indium wires with a diameter of \unit[0.5]{mm} were placed between
the polished surfaces and the stack was compressed with four M4
screws applying a torque of \unit[0.5]{Nm} (see setup shown in
figure~\ref{fig:vacuumSetup}). Indium will create a compression seal
due to its low elasticity modulus  (\unit[10.5]{GPa}
\cite{Werkstoffe}). We proved the tightness of the indium seals with
helium leak tests using a quadrupole mass spec\-tro\-me\-ter and
found the signal to be dominated by the slow rise due to diffusion
through the chamber walls. A diminutive amount of vacuum grease
 is used on all surfaces to be able to
separate the pieces without excessive application of force because
indium is sticky and would otherwise be difficult to
remove when reopening the chamber.

\subsection{Vacuum system}
\label{sec:Vacuumsystem}

The vacuum parts visible in figure~\ref{fig:setupTopView} connect to
further standard stainless steel UHV components. The main components
are an ion getter pump, a titanium sublimation pump and an ion
gauge, all connected via standard CF40 parts.

All steel parts were baked on air at approximately
\unit[300]{$^{\circ}$C} for three days, to reduce the achievable
vacuum pressure \cite{Park2008}. In a second step, we connected the
setup shown in figure~\ref{fig:vacuumSetup} to a turbo molecular
pump and applied a second clean bake at approximately
\unit[120]{$^{\circ}$C} (compatible with all components built in and
below the melting temperature of indium at \unit[156.4]{$^\circ$C}
\cite{Werkstoffe}) for three days to remove any excess amount of
vacuum grease from the system. After this cleaning bake, the ion
getter pump, the titanium sublimation pump and the ion gauge were
assembled and the final bake at approximately
\unit[120]{$^{\circ}$C} was applied for four days.

After four
 months of daily ion trapping the pressure is in the low \unit[$10^{-11}$]{mbar} range.
We can observe
\ybg-ions over a time span of approximately seven hours. This was estimated from
trapping of larger samples and observing the lifetime of the complete chain.

\section{Electrical Interconnect}
\label{sec:ElectricalVacuumInterconnect}

\begin{figure}
\centering
\includegraphics[width=0.95\columnwidth]{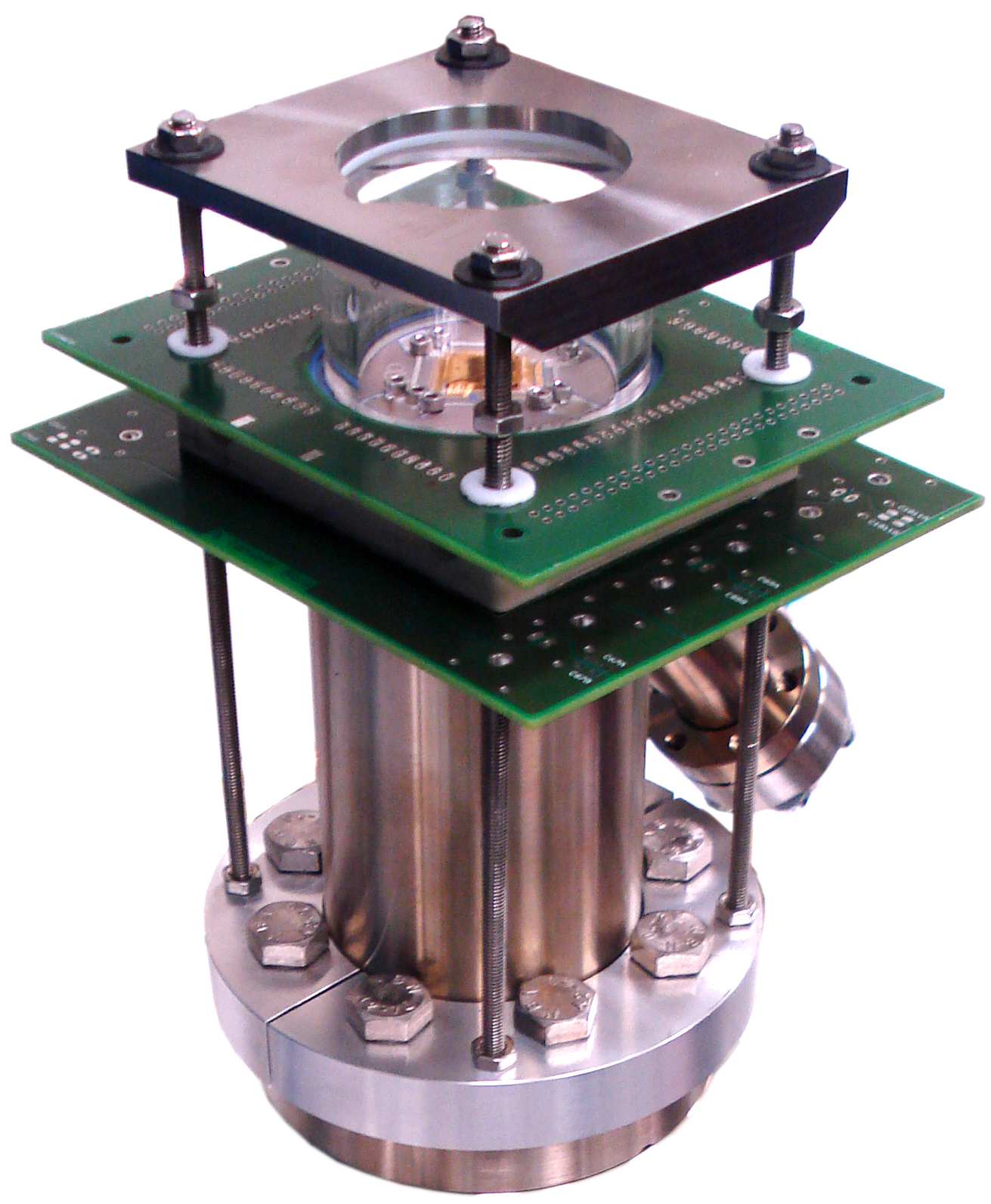}
\caption{ Essential vacuum setup: a simple and compact custom made
steel chamber with a CF~50 flange provides the basis for the chip
holder, and provides an exit port for the laser beams through a
CF~16 flange view port. On the polished surface of the tube the
ceramic chip carrier is placed, and on top of that a cylindrical
glass cap with an interferometric quality window, anti-reflection
coated for all our laser wavelengths. The chip carrier is connected
to printed circuit boards (see
section~\ref{sec:ElectricalVacuumInterconnect}). The multipin
connectors are not shown. The whole stack is compressed with a steel
frame by threaded rods.}\label{fig:vacuumSetup}
\end{figure}

All control voltages and currents are supplied to the microtrap carrier via two double-layer printed circuit boards (PCB)
which provide quick and convenient multi-pin connectors to connect to any signal source (see figure~\ref{fig:vacuumSetup}).
 The high thermal conductivity and the large mass of the chip
 carrier (compared to the PCBs) exclude the use of conventional
 soldering techniques (soldering and reflow soldering). As an
 alternative, conductive epoxy is used to connect
 the chip carrier to the PCBs.

\begin{figure}[t]
    \centering
        \includegraphics[width=0.5\columnwidth]{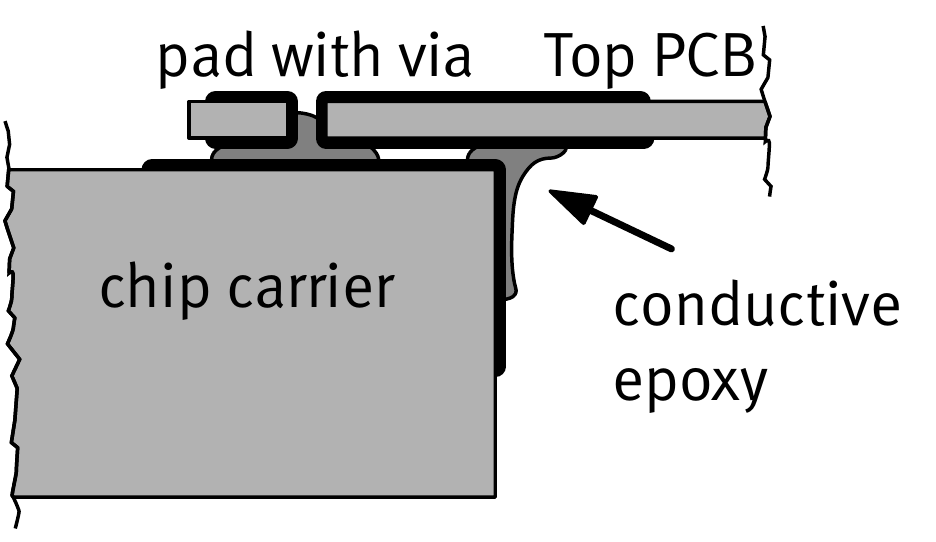}
\caption{Glueing of the microtrap carrier and the top-PCB: a first
droplet of conductive epoxy is applied through the via in the pad
using a syringe after checking for correct alignment. A second
droplet applied on the side strengthens the mechanical connection
and improves the resistance of the contact. A gap of \unit[0.3]{mm}
allows to control the spreading of glue between the chip carrier and
the top-PCB.} \label{fig:Top-PCB-Glueing}
\end{figure}

The upper PCB (in the following called top-PCB) is connected to the
chip carrier as indicated in figure~\ref{fig:Top-PCB-Glueing}. The
large number of pads on this board requires small sizes and
increases the risk of accidental shortcuts from spreading glue. To
resolve this problem, each pad on the top-PCB has a
\unit[0.95]{mm}-hole, which can be used to check the alignment (to
the corresponding pad on the chip carrier) and to insert the correct
amount of glue using a syringe. To allow for controlled spreading of
the glue we found that a gap of approximately \unit [0.3]{mm}
between chip carrier and PCB is essential.
An additional drop of
epoxy at the side of the carrier (see
figure~\ref{fig:Top-PCB-Glueing}) increases the mechanical strength
and conductivity of the connection. This technique ensures fault
free connectivity for all connections of top-PCB by providing the
possibility to check and repair (if required) any bad contact or
short cut between neighbouring pads before proceeding to curing. The resistance between bond pads and the top-PCB after
curing was found to be below \unit [15]{\textohm} for all
connections.

All dc electrodes require currents only in the mA-regime, whereas
higher currents are necessary to drive our atom ovens (up to
\unit[2.5]{A}, see section~\ref{sec:trapping}) or to generate
magnetic fields gradients. All higher currents are provided through the
lower board (called bottom-PCB). We achieved substantially lower
resistivity compared to the upper side using tapered wires on the
chip carrier and large pad areas (\unit [65]{mm$^2$} -- \unit
[121]{mm$^2$}) for large cross sections of the conductive epoxy
connections, accompanied by thick, short and wide traces on the
bottom-PCB. On the other hand, scaling up the width per connection
results in a much lower number of connections on this board (11
total). The resistance between chip carrier and the bottom-PCB after
curing was found to be below \unit [90]{m\textohm} for all
connections.

\section{Trapping potentials}
\label{sec:TrappingPotentials}

The rf potential required for the Paul trap is provided to the
microtrap using a frequency generator which can be amplified before
feeding the signal to a helical resonator (for design details, see
\cite{Weisgerber2003,Macalpine1959,Vizmuller1995,Zverev1967}).
Apart from designing the trap for low capacitance,
we tried to keep the load capacitance small (about \unit
[30]{pF}) by connecting the helix to the microtrap with a short
(\unit [5]{cm}) co-axial cable and short connections on the top-PCB
and the chip carrier. The rf feedthrough introduces a capacitance of about \unit[1-3]{pF}. The $Q$-factor of
the helix with connected microtrap system exceeds 500, which is
comparable to the $Q$-factor obtained with a conventionally connected trap.

All dc potentials for the segmented electrodes are provided by
computer-controlled DAC channels  via a routing device which allows
 each signal to be connected to any number or pattern of electrodes.

We also developed an Electric Field Generator (EFG)
\cite{EFG-patent} which provides versatile dc potentials required for a fast
and detailed control of the axial trapping potential, e.g. for
shuttling of ions between trapping and processing zones. The EFG in
its present version provides 20 channels and is planned to be
extended to 100 channels in the future. Potentials up to
\unit[15]{\Vpp} can be controlled with an update rate up to
\unit[20]{MHz}, a precision of approximately \unit[1]{mV}, and a
noise level below \unit[$\pm$0.5]{mV}. The device allows to program,
store and quickly recall up to 16 different output waveform patterns
for each channel. High bandwidth and update rates, as possible with this device, are interesting for ultra-fast transport between trap segments \cite{Schulz2006}, or, provided the update rate substantially exceeds the trap drive frequency, even to compensate micromotion where this is impossible with dc (e.g. due to rf field components along the trap axis). For quantum simulations in surface traps, such a device could provide arbitrary superpositions of dc and rf on each pixel electrode, giving full freedom for the lattice geometry of the trapped ions. Details on the EFG will be published elsewhere.

It is desirable to remove stray fields and rf crosstalk on all dc
electrodes to obtain clean trapping potentials. We achieve this
using the printed capacitance on the chip carrier (explained in
section~\ref{sec:ChipCarrier}, and shown in
figure~\ref{fig:printingSteps}), and an additional low-pass filter
for each dc electrode. Each of these filters is similar to a $\pi$-filter with
two capacitors (\unit[1]{\textmu F} each) but with a resistor (\unit
[15]{k\textohm}) instead of an inductor. The filters are implemented
on an exchangable 4-layer PCB. This filter is plugged into the
Top-PCB from the air side, and can easily be replaced, if desired,
by another filter-PCB with a different characteristics such as a
notch filter.

\section{Lasers and detection}
\label{sec:LasersAndDetection}

The fine structure of ytterbium requires two lasers for laser
cooling \cite{Balzer2006}: one laser at \unit[369]{nm} drives the
\Shalf-\Phalf-transition for cooling and detection, a second laser
at \unit[935]{nm} drives the \Dhalf-\jK-transition to avoid optical
pumping into the metastable \Dhalf state. Another laser at
\unit[638]{nm} is useful to pump back atoms from the \Fstate state
which is occasionally populated by collisions. A fourth laser at
\unit[399]{nm} drives the \Szero-\Pone-transition in neutral
ytterbium and is used for isotope selective photoionization
\cite{Johanning2011}. A feature of ytterbium is the availability of
all these wavelengths from laser diodes. All lasers are set up as
external cavity diode lasers in Littrow configuration and locked to
pressure sealed temperature stabilized Fabry-Perot cavities by a
side of fringe lock. The \unit[369]{nm} laser diode needs to be
cooled down to \unit[$-5$]{$^\circ$C} to reduce the free-running
wavelength \cite{Kielpinski2006} by \unit[1.9]{nm} and requires an
air tight housing to avoid condensation of water on the cold
surface. The lasers are coupled into single mode fibres and are
combined with dichroic mirrors into a single beam which is focused
by a beam shaping telescope into the trap
 region. The fibre collimators are adjusted to compensate chromatic aberrations of the focussing telescope.
The beam is directed with two mirrors through the trap slit.
One of these mirrors is a piezo motor controlled mount with an alignment precision of sub
micrometer and it determines the position where the laser beams cross the trap.

The fluorescent light from the trapped ions is collimated by a
custom made lens system \cite{Schneider2007} (see also
figure~\ref{fig:setupTopView}), optimized for a large numerical
aperture of 0.4, whilst maintaining a diffraction limited spot over
a field of view of \unit[500]{\textmu m}.
The collimated resonance fluorescence can be directed either to a photo multiplier tube (PMT) 
or to an EMCCD Camera. Simultaneous observation of resonance
fluorescence from ions on the camera and from neutral atoms on the
PMT is also possible using a dichroic mirror. We use a single-band
pass filter with a transmission peak at \unit[370]{nm} and a width
(FWHM) of \unit[6]{nm} in front of the camera to detect fluorescence
light from the ion. Additional adjustable apertures are placed in
auxiliary focal planes to further discriminate against unavoidable
stray-light from the trap electrodes at the cooling wavelength of
\unit[369]{nm}.

\section{Loading and trapping}
\label{sec:trapping}

To start trapping ions, all lasers are locked to temperature
stabilized cavities and set to the correct wavelength
\cite{McLoughlin2011}. The reproducibility of our home-built
scanning Michelson lambdameter and all lasers is on the order of a
few tens of Megahertz and sufficient to ensure successful loading,
once the correct wavelengths are set.

We create collimated beams of neutral ytter\-bium at\-oms by ohmic
heating of a small sample of granular ytterbium in a miniature
tube-shaped steel oven (for design, modelling and operation, see
\cite {Johanning2011,Eiteneuer2009}). Two identical ovens, mounted
to the oven holder (see figure~\ref{fig:setupTopView}) approximately
aligned in a plane normal to the laser beam, are filled with
isotopically enriched samples of $^{171}$Yb and $^{172}$Yb,
respectively. Deposition of ytterbium atoms on electrodes in the
regions relevant for high fidelity experiments is avoided by
skimming the atomic beams by apertures. Further reduction of neutral
 atom deposition on trap electrodes is accomplished  by aligning the
 atomic beam such that it intersects the trap axis
approximately in the middle of the wide loading region. At this position, due to the
separation of approximately \unit[500]{\textmu m} from the taper region, the influence
of unwanted axial rf field components is sufficiently screened to be negligible.

Atomic ions are produced using a two-step photo-ionization process, with a resonant
first step 
(see \cite{Johanning2011} for details). When there is no filter in
the detection channel discriminating light at \unit[399]{nm}, the
effectiveness of photoionization can be deduced from the neutral
atom fluorescence. Visible resonance fluorescence from neutral atoms
can serve as a first frequency reference and calibration of the
lambdameter with an atomic standard. It also indicates excessive
neutral atom flux, which is detrimental for our vacuum and causes
undesirable coating of the electrodes with ytterbium.

\begin{figure}[t]
    \centering
        \includegraphics[width=0.95\columnwidth]{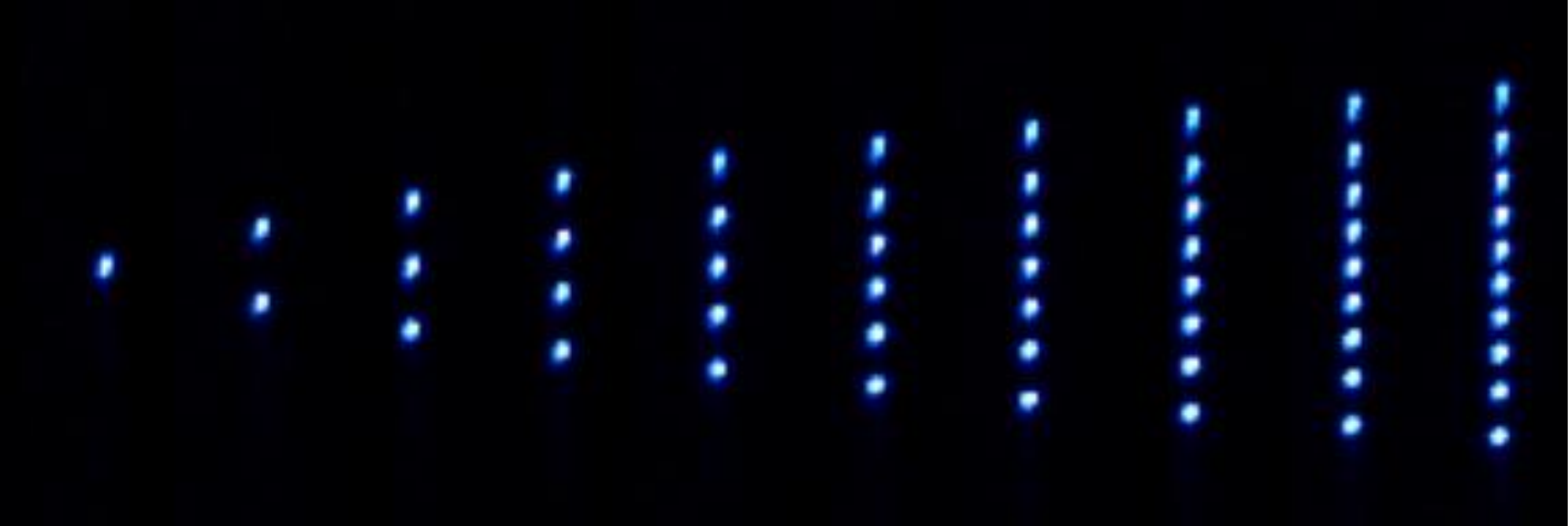}
\caption{Loading of an ion chain: above we show a sequence of
spatial resolved resonance fluorescence images recorded by the EMCCD
camera (false color coded). Each picture (from left to right) with a
string of ions  represents a time-step, where one additional ion was
loaded. All pictures were taken in a single loading run. The
separation between two ions was determined to approximately
\unit[17]{\textmu m}, corresponding to an axial trap frequency of
$\nu_z \approx 2\pi\cdot$~\unit[95]{kHz}.} \label{fig:loaDChain}
\end{figure}

We usually set the oven current, such that the ion loading rate is
small enough (less than one per second) to load ions one by one. The
first ion is loaded during the warm-up of the oven after
approximately two minutes, starting from a cold oven. The loading
rate is then slowly accelerated and longer chains can be loaded within
a few minutes (see figure~\ref{fig:loaDChain}). Applying an rf power
of \unit[125]{mW} results in a radial trapping
frequency of \unit[888]{kHz}, corresponding to an rf amplitude of
approximately \unit[200]{\Vpp} and a stability parameter of $q
\approx 0.22$. The rf drive frequency is
\unit[7.7]{MHz}. The pressure rises during the loading process up
to the mid-\unit[10$^{-11}$]{mbar} range.

\section{Outlook}
\label{sec:Outlook} We showed that thick-film technology allows to
generate customized vacuum feed\-throughs for electrical signals.
These feed\-throughs were shown to be suitable for dc and rf signals
with voltages up to a few hundreds of volts and for currents up to a
few amperes. Gold pastes are available with up to eight times higher
conductivity and should allow for even smaller current interfaces or
larger currents. Moreover, it should be possible, using a proper
design, to extend the frequency range into the microwave regime.
Thick-film interfaces allow for compact designs with very good
access for lasers, detection, rf- and microwave-fields. Such
interfaces are useful in general atomic and molecular physics,
especially when working with micro-structured components, such as
traps for neutral atoms or ions. The vacuum compatibility was found
to be satisfying and matches usual requirements in atomic and
molecular physics, as we regularly attain pressures in the low
\unit[$10^{-11}$]{mbar} range.

We also used similar multi-layered thick-film based chip carriers as
pure in-vacuum devices as they allow for complex structures, as, for
example, solder free filter boards in the immediate vicinity of the
trap to passively remove noise from dc electrodes. For this purpose
not only conductive and isolating pastes are available (which alone
are sufficient to create capacitors) but also resistive pastes for
printed resistors and dedicated dielectric pastes to create compact
capacitors. The properties of such circuits can be fine-tuned
precisely by changing the area of printed structures using laser
trimming \cite{Gupta2005}. Even more applications are possible by
thick-film hybrid technology, which combines thick-film technology
with conventional (glued or soldered) SMD elements.

The scope of the experiment described in this article is to utilize
switchable and tunable magnetic field gradients to use MAGIC.
Together with the flexibility of the axial trapping potential
\cite{McHugh2005,Wunderlich_H2009,Schulz2006} this should allow for
interesting experiments rooted in quantum information science and to
generate tailored Hamiltonians for quantum simulations.

The strong magnetic field gradient extends over a few hundreds of
micrometers and ions need to be shuttled into the experiment region
(located approximately in the middle of the narrow trap slit, see
figure~\ref{fig:BondedChip}) to experience its effect. Thereupon
MAGIC can be utilized using radio frequency fields operating on
Zeeman levels of the \Shalf ground state or, for an easier shelving
on Zeeman levels of the \Dhalf manifold \cite{Johanning2009b}.
Another possibility is to use the isotope \ybuion which provides a
nuclear spin of $I=1/2$ and is thus more difficult to laser cool,
since the hyperfine structure requires additional repumping fields.

MAGIC provides the potential to perform quantum information science
experiments using Doppler cooled ions, removing the obstacle (or
nuisance) of ground state cooling, associated with many
implementations of ion trap based quantum logic. But on the other
hand, as the gradient couples internal and motional states, it
allows for sideband cooling using microwaves, simultaneous cooling
of many vibrational modes \cite{Wunderlich2005} and allows to
improve laser cooling altogether \cite{Albrecht2011}.

\section*{Appendix A: Technical Details}

The microtrap chip carrier consists of a \unit[15]{mm} thick block
of alumina with outer dimensions of \unit[65]{mm} $\times$
\unit[75]{mm}. It was milled by MicroCeram GmbH Meissen. The
thick film printing of the block was carried out by the Hybrid Laboratory
of the University of Siegen.

For the thick film prints, isolating paste Du\-Pont 5704, gold paste
Du\-Pont 5722L and silver-palladium paste Du\-Pont LF 121 were used.
As the thickness of the alumina block is about 15 times higher than
typical thick-film printed ceramics, the firing process had to be
adjusted. Instead of a standard \unit[50]{K/min} slope a maximum
velocity of \unit[20]{K/min} was implemented. The maximum
temperature of \unit[850]{$^\circ$C} was held for \unit[20]{min}
instead the typical \unit[10]{min}. The whole cycle lasted
\unit[120]{min}. The typical height of printed thick film structures
is \unit[7-11]{\textmu m}.

For the polishing procedure,
polishing heads with a ring shaped end face
were built of aluminium to be able to just polish the ring shaped areas of
isolation paste described above. The polishing heads were equipped
with special cloth for the polishing pastes, Winter-Diamant
polishing cloth D0,7-D0,25, of Saint Gobain Diamant\-werk\-zeu\-ge.
The polishing heads were then mounted on a drilling machine, operating at 480 rounds per
minute, and the isolation rings were polished for about \unit[10]{s} with diamond-pastes
Winter-Diaplast N~D~0,7 with a particle size of \unit[0.7]{\textmu m} or N~D~0,25 with a particle size of \\unit[0,25]{\textmu m} with small force.

After this, the surface was checked under a light cut microscope.
Typically after two or three polishing cycles, the surface no longer
showed the screen printing structure and was smoothly polished.
On such a surface, the indium seals were always tight.

For the sealing, indium wire with a diameter of \unit[0.5]{mm} and a
pureness of \unit[99.99]{\%} is used. Before putting the indium wire onto the
surfaces, a diminutive amount of vacuum grease Pfeiffer Vacuum
BN845805-T  is applied. After that, the indium was formed to a ring
of a diameter according to the diameter of the isolating rings. The
endings of the indium wire were crossed over each other, before
putting the indium ring onto the surface.

The pastes used in thick-film technology are hardening slowly during
storage and processing and the viscosity rises perceptibly. We
believe this is the reason that in some of our prints, we found
small leaks along a few radial wires when leak testing with helium,
presumably caused by micro channels. As a precaution, we applied a
narrow layer of leak sealing Torr Seal Vacuum Epoxy, Varian, along
the edges of the isolating pastes to seal all potential channels and
found that this procedure results in  helium tight interfaces
reliably. We expect this sealing step to be unnecessary, when using
fresh paste with low viscosity for thick-film printing.

Detailed design guidelines specific for the thick-film-printing
facilities at Siegen are available upon request from the authors.

\section*{Acknowledgments}
We gratefully acknowledge our electrical and mechanical work shops
and especially D.~Gebauer, who accomplished the thick-film printing
for our chip carrier. We thank Andr\'{e}s F. Var\'{o}n for fruitful
discussions. We acknowledge financial support by the European Union
(STREP Microtrap and PICC), by the Deut\-sche
For\-schungs\-ge\-mein\-schaft and by secunet AG.

\bibliographystyle{revtex}
\bibliography{mutrap1stTrapping}

\end{document}